\documentclass[10pt,conference]{IEEEtran}

\ifCLASSOPTIONcompsoc
  \usepackage[nocompress]{cite}
\else
  \usepackage{cite}
\fi

\usepackage{graphicx}

\usepackage{dashrule}

\usepackage[linesnumbered,ruled,vlined]{algorithm2e}
\usepackage{algpseudocode}
\usepackage{amsmath,amssymb}
\usepackage{multirow}
\usepackage{xspace}

\usepackage[numbers, square, comma, sort&compress]{natbib}

\usepackage{url}
\DeclareFixedFont{\ttb}{T1}{txtt}{bx}{n}{10} 
\DeclareFixedFont{\ttm}{T1}{txtt}{m}{n}{10}  

\usepackage{color}
\definecolor{deepblue}{rgb}{0,0,0.5}
\definecolor{deepred}{rgb}{0.6,0,0}
\definecolor{deepgreen}{rgb}{0,0.5,0}

\usepackage{listings}

\newcommand\pythonstyle{\lstset{
		language=Python,
		basicstyle=\ttm,
		otherkeywords={self},             
		keywordstyle=\ttb\color{deepblue},
		emph={MyClass,__init__},          
		emphstyle=\ttb\color{deepred},    
		stringstyle=\color{deepgreen},
		frame=tb,                         
		showstringspaces=false,            %
		numbers=left
}}

\makeatletter
\def\lst@makecaption{%
  \def\@captype{table}%
  \@makecaption
}
\makeatother

\lstnewenvironment{python}[1][]
{
	\pythonstyle
	\lstset{#1}
}
{}

\usepackage[text size=small, backgroundcolor=blue!30,bordercolor=blue,linecolor=blue]{todonotes}

\newcommand{\myparagraph}[1]{\vspace{1.5ex}\noindent\emph{#1}\hspace{1.5ex}}

\newcommand{\substepseparator}{\hspace{1cm}}

\newcommand{\swat}{SWaT\xspace}
\newcommand{\cps}{CPS\xspace}
\newcommand{\plc}{PLC\xspace}
\newcommand{\plcs}{PLCs\xspace}
\newcommand{\ml}{ML\xspace}
\newcommand{\svm}{SVM\xspace}

\begin{document}

\title{Learning from Mutants:\\ Using Code Mutation to Learn and Monitor Invariants of a Cyber-Physical System}

\author{\IEEEauthorblockN{Yuqi Chen, Christopher M. Poskitt, and Jun Sun}
\IEEEauthorblockA{Singapore University of Technology and Design\\
Singapore, Singapore\\
Email: yuqi\_chen@mymail.sutd.edu.sg; \{chris\_poskitt, sunjun\}@sutd.edu.sg}}

\maketitle

\begin{abstract}
	Cyber-physical systems (\cps) consist of sensors, actuators, and controllers all communicating over a network; if any subset becomes compromised, an attacker could cause significant damage. With access to data logs and a model of the \cps, the physical effects of an attack could potentially be detected before any damage is done. Manually building a model that is accurate enough in practice, however, is extremely difficult. In this paper, we propose a novel approach for constructing models of \cps automatically, by applying supervised machine learning to data traces obtained after systematically seeding their software components with faults (``mutants''). We demonstrate the efficacy of this approach on the simulator of a real-world water purification plant, presenting a framework that automatically generates mutants, collects data traces, and learns an SVM-based model. Using cross-validation and statistical model checking, we show that the learnt model characterises an invariant physical property of the system. Furthermore, we demonstrate the usefulness of the invariant by subjecting the system to 55 network and code-modification attacks, and showing that it can detect 85\% of them from the data logs generated at runtime.
\end{abstract}

\IEEEpeerreviewmaketitle

\section{Introduction}\label{sec:introduction}

Cyber-physical systems~(\cps), in which software components and physical processes are deeply intertwined, are found across engineering domains as diverse as aerospace, autonomous vehicles, and medical monitoring; they are also increasingly prevalent in public infrastructure, automating critical operations such as the management of electricity demands in the grid, or the purification of raw water~\cite{Khaitan-McCalley15a,Lee08a}. In such applications, \cps typically consist of distributed software components engaging with physical processes via sensors and actuators, all connected over a network. A compromised software component, sensor, or network has the potential to cause considerable damage by driving the actuators into states that are incompatible with the physical conditions~\cite{Cardenas-Amin-Sastry08a}, motivating research into practical approaches for monitoring and attesting \cps to ensure that they are operating safely and as intended.

Reasoning about the behaviour exhibited by a \cps, however, is very challenging, given the tight integration of algorithmic control in the ``cyber'' part with continuous behaviour in the ``physical'' part~\cite{Zheng-et_al15a}. While the software components are often simple when viewed in isolation, this simplicity betrays the typical complexity of a \cps when taken as a whole: even with domain-specific expertise, manually deriving accurate models of the physical processes (e.g.~ODEs, hybrid automata) can be extremely difficult---if not impossible. This is unfortunate, since with an accurate mathematical model, a supervisory system could query real \cps data traces and determine whether they represent correct or compromised behaviour, raising the alarm for the latter.

In this work, using a high degree of automation, we aim to overcome the challenge of constructing \cps models that are useful for detecting attacks in practice. In particular, we propose to apply machine learning (\ml) on traces of sensor data to construct models that characterise \emph{invariant properties}---conditions that must hold in all states amongst the physical processes controlled by the \cps---and to make those invariants checkable at runtime. In contrast to existing unsupervised approaches (e.g.~\cite{Goh_et-al17a,Inoue-et_al17a}), we propose a supervised approach to learning that trains on traces of sensor data representing ``normal'' runs (the positive case, satisfying the invariant) as well as traces representing abnormal behaviour (the negative case), in order to learn a model that characterises the border between them effectively. To systematically generate the negative traces, we propose the novel application of code mutation (\`{a} la mutation testing~\cite{Jia-Harman11a}) to the software components of \cps. Motivating this approach is the idea that small syntactic changes may correspond more closely to an attacker attempting to be subtle and undetected. Once a \cps model is learnt, we propose to use statistical model checking~\cite{Clarke-Zuliani11a} to ascertain that it is \emph{actually} an invariant of the \cps, allowing for its use in applications such as the physical attestation of software components~\cite{Roth-McMillin13a} or runtime monitoring for attacks.

In order to evaluate this approach, we apply it to Secure Water Treatment~(\swat)~\cite{SWaT-Reference}, a scaled-down but fully operational water treatment testbed at the Singapore University of Technology and Design, capable of producing five gallons of safe drinking water per minute. \swat has industry-standard control software across its six Programmable Logic Controllers (\plcs). While the software is structurally simple, it must interact with physical processes that are very difficult to reason about, since they are governed by laws of physics concerning the dynamics of water flow, the evolution of pH values, and the various chemical processes associated with treating raw water. In this paper, we focus on water flow: we learn invariants characterising how water tank levels evolve over time, and show their usefulness in detecting both manipulations of the control software (i.e.~attestation) as well as detecting attacks in the network that manipulate the sensor readings and actuator signals. Our experiments take place on a simulator of \swat due to resource restrictions and safety concerns, but the simulator is faithful and reasonable: it implements the same PLC code as the testbed, and has a cross-validated physical model for water flow.

\myparagraph{Our Contributions.} This paper proposes a novel approach for generating models of \cps, based on the application of supervised machine learning to traces of sensor data obtained after systematically mutating software components. To demonstrate the efficacy of the approach, we present a framework for the \swat simulator that: (1)~automatically generates mutated PLC programs; (2)~automatically generates a large dataset of normal and abnormal traces; and (3)~applies Support Vector Machines~(\svm) to learn a model. We apply cross-validation and statistical model checking to show that the model characterises an invariant physical property of the system. Finally, we demonstrate the usefulness of the invariant in two applications: (1)~code attestation, i.e.~detecting modifications to the control software through their effects on physical processes; and (2)~identifying standard network attacks, in which sensor readings and actuator signals are manipulated.

This work follows from the ideas presented in our earlier position paper~\cite{Chen-Poskitt-Sun16a}, but differs significantly. In particular, the preliminary experiment in~\cite{Chen-Poskitt-Sun16a} was entirely manual, used (insufficiently expressive) linear classifiers, had a very limited dataset, and only briefly discussed how the invariants might be evaluated. In the present paper, we work with significantly larger datasets that are generated automatically, learn much more expressive classifiers using kernel methods, use a systematic approach to feature vector labelling, apply statistical model checking to validate the model, and assess its usefulness for detecting network and code-modification attacks.

\myparagraph{Organisation.}
The remainder of the paper is organised as follows. In Section~\ref{background}, we introduce the \swat water treatment system as our motivating case study, and present a high-level overview of our approach. In Section~\ref{mutantandlearn}, we describe in detail the main steps of our approach, as well as how it is implemented for the \swat simulator. In Section~\ref{evaluation}, we evaluate our approach with respect to five research questions. In Section~\ref{related_work}, we highlight some additional related work. Finally, in Section~\ref{conclusion}, we conclude and suggest some directions for future work.

\section{Motivation and Overview}\label{background}

In this section, we introduce \swat, the water treatment \cps that provides our motivation for learning and monitoring invariants, and also forms the case study for this paper. Following this, we present a high-level overview of our learning approach and how it can be applied to \cps.

\subsection{Motivational Case Study: \swat Testbed}\label{swat_testbed}

\begin{figure}[!t]
	\centering
	\includegraphics[width=\linewidth]{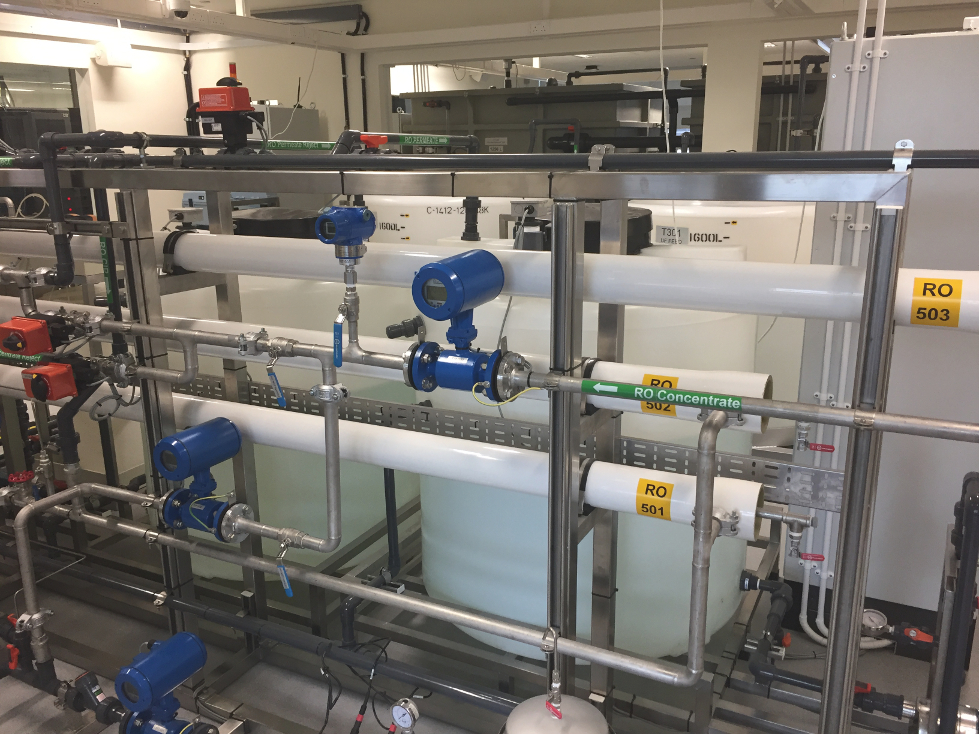}
	\caption{The Secure Water Treatment (\swat) testbed}
	\label{fig:swat_testbed}
\end{figure}

The \cps that forms the case study for our paper is Secure Water Treatment (\swat)~\cite{SWaT-Reference}, a testbed built for cyber-security research at the Singapore University of Technology and Design (Figure~\ref{fig:swat_testbed}). \swat is a scaled-down but fully operational raw water purification plant, capable of producing five gallons of safe drinking water per minute. Raw water is treated in six distinct but co-operating stages, handling chemical processes such as ultrafiltration, de-chlorination, and reverse osmosis.

Each stage of \swat consists of a dedicated programmable logic controller~(\plc), which communicates over a ring network with some sensors and actuators that interact with the physical environment. The sensors and actuators vary from stage-to-stage, but a typical sensor in \swat might read the level of a water tank or the water flow rate in some pipe, whereas a typical actuator might operate a motorised valve (for opening an inflow pipe) or a pump (for emptying a tank). A historian records the sensor readings and actuator signals, facilitating large datasets for offline analyses~\cite{Goh-et_al16a}.

The \plcs are responsible for algorithmic control in the six stages, repeatedly reading sensor data and computing the appropriate signals to send to actuators. The programs that \plcs cycle through every 5ms are structurally simple. They do not contain any loops, for example, and can essentially be viewed as large, nested conditional statements for determining the interactions with the system's 42 sensors and actuators. The programs can easily be viewed (in both a graphical and textual style), modified, and re-deployed to the \plcs using Rockwell's RSLogix 5000, an industry-standard software suite.

In addition to the testbed itself, a \swat simulator~\cite{Framework} was also developed at the Singapore University of Technology and Design. Built in Python, the simulator faithfully simulates the cyber part of \swat, as a direct translation of the \plc programs  was possible. Inevitably, the physical part (taking advantage of Python's scientific libraries, e.g.~NumPy, SciPy) is less accurate since the actual ODEs governing \swat are unknown. The simulator currently models some of the simpler physical processes such as water flow (omitting, for example, models of the chemical processes), the accuracy of which has been improved over time by cross-validating data from the simulator with real \swat data collected by the historian~\cite{Goh-et_al16a}. As a result, the simulator is especially faithful and useful for investigating over- and underflow attacks on the water tanks.

The \swat testbed characterises many of the security concerns that come with the increasing automation of public infrastructure. What happens, for example, if part of the network is compromised and packets can be manipulated; or if a PLC itself is compromised and the control software replaced? If undetected, the system could be driven into a state that causes physical damage, e.g.~activating the pumps of an empty tank, or causing another one to overflow. The problem (which this paper aims to overcome) is that detecting an attack at runtime is very difficult, since a monitor must be able to query live data against a model of how \swat is actually expected to behave, and this model must incorporate the physical processes. As mentioned, the \plc programs in isolation are very simple and amenable to formal analysis, but it is impossible to reason about the system as a whole without incorporating some knowledge of the physical effects of actuators over time.

\subsection{Overview of Our Approach}

Our approach for constructing \cps models consists of three main steps, as sketched in Algorithm~\ref{alg0}: (1)~simulating the \cps under different code mutations to collect a set of normal and abnormal system traces; (2)~constructing feature vectors based on the two sets of traces and learning a classifier; and (3)~applying statistical model checking to determine whether the classifier is an invariant, restarting the process if it is not. In the following we provide a high-level overview of how these three steps are applied in general. A more detailed presentation of the steps and their application to the \swat simulator are given later, in Section~\ref{mutantandlearn}.

\begin{algorithm}[t]
	\label{alg0}
	\caption{Sketch of Overall Algorithm}
	\KwIn{A CPS $S$}
	\KwOut{An invariant $\phi$}
    Randomly simulate $S$ for $n$ times and collect a set of normal traces $Tr$; \\
    Construct a set of mutants $Mu$ from $S$; \\
    Collect a set of positive feature vectors $Po$ from $Tr$; \\
    Collect a set of negative feature vectors $Ne$ based on abnormal traces from $Mu$; \\
    Learn a classifier $\phi$;\\
    Apply statistical model checking to validate $\phi$;\\
    \If {$\phi$ satisfies our stopping criteria} {
        return $\phi$; \\
    }
    Restart with additional data;
\end{algorithm}

In the first step, traces (e.g.~of sensor readings) representing normal system behaviour are obtained by randomly simulating the \cps under normal operating conditions, i.e.~with the cyber part (\plcs) and physical part (ODEs) unaltered. To collect traces representing abnormal behaviour, our approach proposes simulating the \cps under small manipulations. Since we aim for our learnt invariants to be useful in detecting \plc and network attacks (as opposed to attackers tampering directly with the environment), we limit our manipulations to the cyber part, and propose a systematic method motivated by the assumption that attackers would attempt to be subtle in their manipulations. Our approach is inspired by mutation testing~\cite{Jia-Harman11a}, a fault-based testing technique that deliberately seeds errors---small, syntactic transformations called \emph{mutations}---into multiple copies of a program. Mutation testing is typically used to assess the quality of a test suite (i.e.~good suites should detect the mutants), but in our approach, we generate mutants from the original \plc programs, and use these modified instances of the \cps to collect abnormal data traces. 

In the second step, we extract positive and negative feature vectors from the normal and abnormal data traces respectively. Since an attack (i.e.~some modification of a PLC program or a network attack) takes time to affect the physical processes, our feature vectors are pairs of sensor readings taken at fixed time intervals. While feature vectors can be extracted from the normal traces immediately, some pre-processing is required before they can be extracted from the abnormal ones: the mutations in some mutant \plc programs may never have been executed, or only executed after a certain number of system cycles, leading to traces either totally or partially indistinguishable from positive ones. To overcome this, we compare abnormal traces with normal ones obtained from the same initial states, discarding wholly indistinguishable traces, and then extracting pairs of sensor readings only when discrepancies are detected. With the feature vectors collected, we apply a supervised ML algorithm, e.g.~Support Vector Machines (SVM), to learn a classifier.

In the third step, we must validate that the classifier is actually an invariant of the \cps. After applying standard ML cross-validation to minimise generalisation error, we apply statistical model checking~(SMC)~\cite{Clarke-Zuliani11a} to establish whether or not there is statistical evidence that the model is an invariant. In SMC, additional normal traces of the \cps are observed, and statistical estimation or hypothesis testing (e.g.~the sequential probability ratio test~(SPRT)~\cite{Younes-Simmons02a}) is used to estimate the probability of the classifier's correctness. If the probability is high (i.e.~above some predetermined threshold), we take that classifier as our invariant. Otherwise, we repeat the whole process with different randomly sampled data.

With a \cps invariant learnt, a supervisory system can monitor live data from the system and query it against the invariant, raising an alarm when it is not satisfied. This has at least two applications in defending against attacks. First, it can be used to detect standard network attacks, where packets have been manipulated and actuators are shifted into states that are inappropriate for the current physical environment. Second, it can be seen as a form of code attestation: if the actual behaviour of a \cps does not satisfy our mathematical model of it (i.e.~the invariant), then it is possible that the cyber part has been compromised and that ill-intended manipulations are occurring. This form of attestation is known as \emph{physical attestation}~\cite{Roth-McMillin13a,Valente-Barreto-Cardenas14a}, and while weaker than typical software- and hardware-based attestation schemes (e.g.~\cite{Alves-Felton04a,Anati-et_al13a,Castelluccia-et_al09a,Seshadri-et_al04a}), it is much more lightweight as neither the firmware nor the hardware of the PLCs need to be modified.

\section{Implementing Our Approach} \label{mutantandlearn}

In this section, we describe in detail the main steps of our approach: (1)~generating mutants and data traces; (2)~collecting positive and negative feature vectors for learning a classifier; and (3)~validating the classifier.

We illustrate each of the steps in turn by applying them to the \swat simulator. We remark that our choice to use the \swat simulator (rather than the testbed) has some important advantages for this paper. It allows us to automate each step in an experimental framework, with which we can easily investigate the effects of different parameters on the accuracy of learnt models. Furthermore, mutations can be applied and attacks can be simulated without the risk of damage, and the usefulness of learnt invariants can be assessed without wasting resources (e.g.~water, chemicals) or navigating the policy restrictions of the testbed. Obtaining an invariant for the testbed can be achieved by re-running the trace collection phase on \swat with optimised parameters for learning (see Section~\ref{evaluation}), or improving the accuracy of the physical model in the simulator to the extent that learnt classifiers can be validated as invariants of both the simulator and the testbed.

\subsection{First Step: Generating Mutants and Traces}
\label{sec:generating_mutants_and_traces}

The first step of our approach is collecting the traces of raw sensor data that will subsequently be used for learning a \cps invariant. It consists of the following sub-steps: (i)~fixing a set of initial physical configurations and a time interval for taking sensor readings; (ii)~generating data traces that represent normal system behaviour; (iii)~applying mutations and generating the (possibly) abnormal traces they produce.

\myparagraph{Sub-step (i): Initial Configurations.} In order to collect a set of data that captures the \cps' behaviour across a variety of physical contexts, a set of initial configurations should be chosen that covers the extremities of the sensors' ranges, as well as randomly selected combinations of values within them. A time interval for logging sensor readings (e.g.~the historian's default) should also be chosen, as well as a length of time to run the \cps from each initial configuration.

\emph{Applied to \swat.} In the case of the \swat simulator, since it only models physical processes concerning water flow, we collect traces of data from sensors recording the water levels in the five tanks. In particular, physical configurations are expressed in terms of the water levels recorded by these five sensors. The set of initial configurations we use in our experiments (see Section~\ref{evaluation}) therefore includes different combinations of water tank levels, including extreme values (i.e.~tanks being full or empty). We choose to log the sensor values every 5ms, corresponding to the default time interval at which the simulator logs data. We fix 30 minutes as the length of time to run the simulator from each configuration, as previous experimentation has shown that the simulator's physical model remains accurate for at least this length of time.

\substepseparator

\myparagraph{Sub-step (ii): Normal Traces.} To generate normal traces, we simply launch the \cps under normal operating conditions from each initial physical configuration, using the run length and time interval fixed in sub-step~(i). The traces of sensor data should be extracted from the historian for processing in a later step.

\emph{Applied to \swat.} For our case study, we built a framework~\cite{Framework} around the \swat simulator that can automatically launch and run the software on each of the initial configurations chosen earlier. Each run uses the original (i.e.~unaltered) PLC programs, lasts for 30 minutes, and logs the simulated water tank levels every 5ms. These logs are stored as raw text files from which feature vectors are extracted in a later step.

\substepseparator

\myparagraph{Sub-step (iii): Mutants and Abnormal Traces.} Next, we need to generate data traces representing abnormal system behaviour. In order to learn a classifier that is as close to the boundary of normal and abnormal behaviour as possible, we generate these traces after subjecting the control software to small syntactic code changes (i.e.~mutations). These code changes are the result of applying simple mutation operators, which randomly replace some Boolean operator, logical connector, arithmetic function symbol, constant, or variable. To ensure a diverse enough training set, we generate abnormal traces from multiple versions of the control software representing a variety of different mutations.

Our approach for generating mutant PLC programs is summarised in Algorithm~\ref{generating_mutants}. Given a set of co-operating PLC programs, the algorithm makes a copy of all of them, and applies an applicable mutation operator to a single PLC program in the set. 

\begin{algorithm}[!t]
	\label{generating_mutants}
	\caption{Generating Mutant PLC Code}
	\KwIn{A set of PLC programs $S$} 
	\KwOut{A mutant set of PLC programs $S_M$}
	Let $Ops$ be the set of mutation operators; \\
	Make a copy $S_M$ of the PLC programs $S$; \\
	$applied := false$; \\
	\While{$\neg applied $}
	{
		Randomly choose a \plc $P$ from $S_M$; \\
		Randomly choose a line number $i$ from $P$; \\
		\If{\emph{some operator in $Ops$ is applicable to line $i$}}
		{
			Apply an applicable operator to line $i$; \\
			$applied := true$; 
		}

	}
	return $S_M$;	
\end{algorithm}

\begin{figure*}
\noindent\hspace{15pt}\begin{minipage}{.45\textwidth}
\begin{python}[breaklines,caption={Snippet of unmodified control code from PLC \#3},label=lst:unmodified_plc3]%[frame=single,language=Python,breaklines=true]{Name}
if Sec_P:
  MI.Cy_P3.CIP_CLEANING_SEC=HMI.Cy_P3.CIP_CLEANING_SEC+1
  if HMI.Cy_P3.CIP_CLEANING_SEC>HMI.Cy_P3.CIP_CLEANING_SEC_SP or self.Mid_NEXT:
	self.Mid_NEXT=0
	HMI.P3.State=19
break
\end{python}
\end{minipage}\hfill
\begin{minipage}{.45\textwidth}
\begin{python}[breaklines,caption={A possible mutant obtained from Listing~\ref{lst:unmodified_plc3}},label=lst:mutant_plc3]%[frame=single,language=Python,breaklines=true]{Name}
if Sec_P:
  MI.Cy_P3.CIP_CLEANING_SEC=HMI.Cy_P3.CIP_CLEANING_SEC+1
  if HMI.Cy_P3.CIP_CLEANING_SEC>HMI.Cy_P3.CIP_CLEANING_SEC_SP or self.Mid_NEXT:
	self.Mid_NEXT=0
	HMI.P3.State=14
break
\end{python}
\end{minipage}
\end{figure*}

\emph{Applied to \swat.} In the case of the \swat simulator, our framework can automatically and randomly generate multiple mutant simulators. Note that each mutant simulator, built up of six PLC programs, consists of one mutation only in a PLC program chosen at random. Since the PLC programs are syntactically simple, we need only six mutation operators (Table~\ref{mutation_operators}). Evidence suggests that additional mutation operators are unlikely to increase coverage~\cite{Offutt-et_al96a}, so our mutant simulators should be sufficiently varied.

To illustrate, consider the code in Listing~\ref{lst:unmodified_plc3}, a small extract from the PLC program controlling ultrafiltration in \swat. If the guard conditions are met, line 5 will change the state of the PLC to ``19''. This number identifies a branch in a case statement (not listed) that triggers the signals that should be sent to actuators. Now consider Listing~\ref{lst:mutant_plc3}: this PLC program is identical to Listing~\ref{lst:unmodified_plc3}, except for the result of a scalar mutation on line 5 that means the PLC would be set to state ``14'' instead. If executed, different signals will be sent to the actuators, potentially causing abnormal effects on the physical state---as might be the goal of an attacker.

\begin{table}[!t]
	\renewcommand{\arraystretch}{1.3}
	\caption{Mutation Operators}
	\label{mutation_operators}
	\centering
	\begin{tabular}{|c|c|}
		\hline
	  Mutation Operator & Example \\
		\hline
		\footnotesize Scalar Variable Replacement  & $ x=a\ \leadsto\ x=b  $\\
		Arithmetic Operator Replacement & $a+b\ \leadsto\ a-b$ \\
		Relational Operator Replacement & $a>b\ \leadsto\ a\geq b$ \\
		Guard Valuation Replacement  & $\mathtt{if (}c\mathtt{)}  \ \leadsto\ \mathtt{if  (false)} $ \\
		Logical Connector Replacement & $a\ \mathtt{and}\ b \ \leadsto\ a\ \mathtt{or}\ b $ \\
		Assignment Operator Replacement & $ x=a\ \leadsto\ x\ \mathtt{+}\!=a $ \\
		\hline
	\end{tabular}
\end{table}

Once we have our mutant simulators, we discard any that cannot be compiled. Of the mutants remaining, we run them with respect to each initial state for 30 minutes, logging the levels of all the water tanks every 5ms.

The current implementation of our mutant simulator generator for \swat is available online~\cite{Framework}, consisting of just over 200 lines of Python code. It applies mutations to the PLC programs by reading them as text files, randomly choosing a line, and then randomly applying an applicable mutation operator (Table~\ref{mutation_operators}) by matching and substituting. This takes a negligible amount of time, so hundreds of mutant simulators can be generated very quickly (i.e.~in seconds).

\subsection{Second Step: Collecting Feature Vectors, Learning}
\label{sec:collecting_feature_vectors}

At this point, we have a collection of raw data traces generated by normal PLC programs as well as by multiple mutant PLC programs. The second step is to extract positive and negative feature vectors from this data to perform supervised learning. It consists of the following sub-steps: (i) fixing a feature vector type; (ii) collecting feature vectors from the data, undersampling the abnormal data to maintain balance; (iii) applying a supervised learning algorithm.

\myparagraph{Sub-step (i): Feature Vector Type.} A feature vector type must be defined that appropriately represents objects of the data. For traces of sensor data, a simple feature vector would consist of the sensor values at any given time point. For typical \cps however, such a feature vector is far too simple, since it does not encapsulate any information about how the values evolve over the time series---an intrinsic part of the physical model. A more useful feature vector would record the values at fixed time intervals, making it possible to learn patterns about how the levels of tanks change over the time series.

\emph{Applied to \swat.} In the case of the \swat simulator, we define our feature vectors to be of the form $(\pi, \pi')$, where $\pi$ denotes the water tank levels at a certain time and $\pi'$ denotes the values of the same tanks after $d$ time units, where $d$ is some fixed time interval that is a multiple of the interval at which data is logged (we compare the effects of different values of $d$ in Section~\ref{sec:how_large_time_interval}). Our feature vectors are based on the sliding window method that is commonly used for time series data~\cite{Dietterich02a}.

\substepseparator

\myparagraph{Sub-step (ii): Collecting Feature Vectors.} Next, the raw normal and abnormal data traces must be organised into positive and negative feature vectors of the type chosen in sub-step~(i). Extracting positive feature vectors from the normal data is straightforward, but for negative feature vectors, we have the additional difficulty that mutants are not guaranteed to be \emph{effective}, i.e.~able to produce data traces distinguishable from normal ones. Furthermore, even effective mutants may not cause an immediate change. It is crucial not to mislabel normal data as abnormal---additional filtering is required.

\emph{Applied to \swat.} Algorithm~\ref{collecting_feature_vector} summarises how feature vectors are collected from the \swat simulator and its mutants. Collecting positive feature vectors is very simple: all possible pairs of physical states $(\pi, \pi')$ are extracted from the normal traces. For each pair $(\pi, \pi')$ extracted from the abnormal traces, the unmodified simulator is run on $\pi$ for $d$ time units: if the unmodified simulator leads to a state distinguishable from $\pi'$, the original pair is collected as a negative feature vector; if it leads to a state that is indistinguishable from $\pi'$, it is discarded (since the mutation had no effect). In the case of \swat, its simulator is deterministic, allowing for this judgement to be made easily. (For data from the testbed, some acceptable level of tolerance would need to be defined.)

 \begin{algorithm}[!t]
	\label{collecting_feature_vector}
	\caption{Collecting Feature Vectors}
	\KwIn{Set of normal traces $T_N$ and abnormal traces $T_A$, each trace of uniform size $N$} 
	\KwOut{Set of positive feature vectors $Po$; set of negative feature vectors $Ne$}
	Let $S$ be the unmodified simulator;\\
	Let $t$ be the time interval for logging data in traces;\\
	Let $d$ be the time interval for feature vectors;\\
	$x := 0$; $Po := \emptyset$; $Ne := \emptyset$; \\
	\ForEach {$Tr \in T_N$}
	{
	\While{$x + (d/t) < N$}
	{
		$\pi := \langle s_0, s_1, \dots \rangle$ for all sensor values $s_i$ at row $x$ of $Tr$;\\
		$\pi' := \langle s'_0, s'_1, \dots \rangle$ for all sensor values $s'_i$ at row $x+(d/t)$ of $Tr$;\\
		$Po:=Po \cup \{(\pi, \pi')\}$\\
		$x:=x+1$;\\		
	}}
	$x := 0$; \\
	\ForEach {$Tr \in T_A$}
	{
	\While{$x+(d/t) < N$}
	{
		$\pi := \langle s_0, s_1, \dots \rangle$ for all sensor values $s_i$ at row $x$ of $Tr$;\\
		$\pi' := \langle s'_0, s'_1, \dots \rangle$ for all sensor values $s'_i$ at row $x+(d/t)$ of $Tr$;\\
		Run simulator $S$ on configuration $\pi$ for $d$ time units to yield trace $Tr'$;\\
		$\pi'' := \langle s''_0, s''_1, \dots \rangle$ for all sensor values $s''_i$ at row $d/t$ of $Tr'$;\\
		\If{\emph{$\pi' \ne \pi''$}}
		{
			$Ne:=Ne \cup \{(\pi, \pi')\}$
		}
		$x := x+ 1$;\\
	}}
	return $Po$, $Ne$;	
\end{algorithm}

\substepseparator

\myparagraph{Sub-step (iii): Learning.} Once the feature vectors are collected, a supervised ML algorithm can be applied to learn a model.

\emph{Applied to \swat.} For the \swat simulator, we choose to apply SVM as our supervised ML approach since it is fully automatic, with well-developed active learning strategies, and good library support (we use LIBSVM~\cite{Chang-Lin11a}). Furthermore, SVM has expressive kernels and has often been successfully applied to time series prediction~\cite{Sapankevych-Sankar09a}. Based on the training data, SVM attempts to learn the (unknown) boundary that separates it. Different classification functions exist for expressing this boundary, ranging from ones that attempt to find a simple linear separation between the data, to non-linear solutions based on RBF (we compare different classification functions for \swat in Section~\ref{sec:eval_classification_functions}). For the purpose of validating the classifier and assessing its generalisability, it is important to train it on only a portion of the feature vectors, reserving a portion of the data for testing. We randomly select $70\%$ of the feature vectors to use as the training set, reserving the rest for evaluation. 

We remark that SVM can struggle to learn a reasonable classifier if the data is very unbalanced. This is the case for the \swat simulator: we have just one simulator for normal data, but potentially infinite mutant simulators for generating abnormal data. To ensure balance, we undersample the negative feature vectors. Let $N_{Po}$ denote the number of positive feature vectors and $N_{Ne}$ the number of negative feature vectors we collected. We partition the negative feature vectors into subsets of size $N_{Ne} / N_{Po}$ (rounded up to the nearest integer), and randomly select a feature vector from each one. This leads to an undersampled set of negative feature vectors that is roughly the same size as the positive feature vector set.

\subsection{Third Step: Validating the Classifier}
\label{sec:validating_classifier}

At this point, we have collected normal and abnormal data, processed it into positive and negative feature vectors, and learnt a classifier by applying a supervised ML approach. This final step is to determine whether or not there is evidence that the learnt model can be considered a physical invariant of the CPS. It consists of the following two sub-steps: (i) applying standard ML cross-validation to assess how well the classifier generalises; and (ii) apply SMC to determine whether or not there is statistical evidence that the classifier does indeed characterise an invariant property of the system.

\myparagraph{Sub-step (i): Cross-Validation.} Our first validation method is to apply standard ML $k$-fold cross validation (with e.g.~$k=5$) to assess how well the classifier generalises. This technique computes the average accuracy of $k$ different classifiers, each obtained by partitioning the training set into $k$ segments, training on $k-1$, and validating on the segment remaining (repeating with respect to different validation partitions).

\substepseparator

\myparagraph{Sub-step (ii): Statistical Model Checking.} The second validation method applies SMC, a standard technique for verifying general stochastic systems~\cite{Clarke-Zuliani11a}. The variant we use observes executions of the system (i.e.~traces of sensor data), and applies hypothesis testing to determine whether or not the executions provide statistical evidence of the learnt model being an invariant of the system. SMC estimates the probability of correctness rather than guaranteeing it outright. It is simple to apply, since it only requires that we can execute the (unmodified) system and collect data traces. It treats the system as a black box, and thus does not require a model~\cite{Sen-Viswanathan-Agha04a}.

Given some classifier $\phi$ for a system $S$, we apply SMC to determine whether or not $\phi$ is an \emph{invariant} of $S$ with a probability greater or equal to some threshold $\theta$, i.e.~whether $\phi$ correctly classifies the traces of $S$ as normal with a probability greater than $\theta$. Note that the \emph{usefulness} of invariants is a separate question, addressed in Section~\ref{sec:eval_usefulness}. A classifier that always labels normal and abnormal data as normal, for example, is an invariant, but not a useful one for detecting attacks.

\emph{Applied to \swat.} In the case of the \swat simulator, we generate a normal data trace from a new, distinct initial configuration, and collect the positive feature vectors from it. Next, we randomly sample feature vectors from this set, evaluate them with our classifier, and apply SPRT as our hypothesis test to determine whether or not there is statistical evidence that the classifier labels them correctly (setting the error bounds at a standard level of 0.05) with accuracy greater than some $\theta$. If further data is required, we sample additional positive feature vectors from another distinct initial configuration. We remark that we choose $\theta$ to be the accuracy of the best classifier we train in our evaluation (Section~\ref{sec:eval_invariant}). These steps are repeated several times, each with data from additional new initial configurations.

\section{Evaluation} \label{evaluation}

We evaluate our approach through experiments intended to answer the following research questions (RQs):

\begin{itemize}
    \item RQ1: What kind of classification function do we need? 
    \item RQ2: How large should the time interval in feature vectors be?
    \item RQ3: How many mutants do we need?
	\item RQ4: Is our model a physical invariant of the system?
	\item RQ5: Is our model useful for detecting attacks?
\end{itemize}

\noindent RQ1--3 consider the effects of different parameters on the performance of our learnt models, in particular, the classification function (linear, polynomial, or RBF), the different time intervals for constructing feature vectors, and the number of mutants to collect abnormal traces from. We take the best classifier from these experiments, and assess for RQ4 whether or not there is statistical evidence that the model characterises an invariant of the system. Finally, for RQ5, we investigate whether or not the model is useful for detecting various different attacks that manipulate the network and PLC programs.

All the experiments in the following were performed on the \swat simulator~\cite{Framework}. The mutation and learning framework we built for this simulator (as described in Section~\ref{mutantandlearn}) is available to download~\cite{Framework}, and uses version 3.22 of LIBSVM~\cite{Chang-Lin11a} to apply SVM to our feature vectors\footnote{Additional rounds of experiments on different mutants were performed post-publication. The results~\cite{Framework} are consistent with our conclusions here.}.

\subsection{RQ1: What kind of classification function do we need?}
\label{sec:eval_classification_functions}

Our first experiment is to determine which of the main SVM-based classification functions---linear, polynomial (degree 3), or RBF---we should use in order to learn models with an acceptable level of accuracy. Intuitively, a simple model is more useful for human interpretation, but it may not be expressive enough to achieve high classification accuracy. First, we generate 700 mutant simulators, of which 91 are effective (i.e.~led to some abnormal behaviour). From 20 initial configurations of the \swat simulator, as described in Section~\ref{sec:generating_mutants_and_traces}, we generate 30 minute traces (at 5ms intervals) of normal and abnormal data from the original simulator and mutant simulators respectively. From these data traces, we collect $1.68\ast 10^6$ feature vectors with a 250ms time interval type, using undersampling to account for the larger quantity of abnormal data (see Section~\ref{sec:collecting_feature_vectors}). These vectors are then randomly divided into two parts: $70\%$ for training, and $30\%$ for testing. SVM is applied to the training vectors to learn three separate linear, polynomial, and RBF classifiers.

\begin{table*}[!t]
	\renewcommand{\arraystretch}{1.3}
	\caption{Comparison of classification functions}
	\label{tab:comparison_classifiers}
	\centering
\begin{tabular}{c|c|c|c|c}
type &  accuracy & cross-validation accuracy & sensitivity & specificity  \\
\hline
SVM-linear &$63.34\%$  &$64.12\%$ &$66.44\%$ & $60.23\%$  \\
\hline
SVM-polynomial & $67.10\%$&$68.32\%$ &$74.92\%$ & $51.67\%$   \\ 
\hline
SVM-RBF &$91.05\%$ & $90.99\%$&$99.28\%$ & $82.82\%$   \\ 
\end{tabular}
\end{table*}

Table~\ref{tab:comparison_classifiers} presents a comparison between the three classifiers learnt in the experiment. We report two types of accuracy. The \emph{accuracy} column reports how many of the held-out feature vectors (i.e.~the $30\%$ of the collected feature vectors held out for testing) are labelled correctly by the classifier. The \emph{cross-validation accuracy} is the result of applying $k$-fold cross-validation (with $k=5$) to the training set: this is the average accuracy of five different classifiers, each obtained by partitioning the training set into five, training on four partitions, and validating on the fifth (then repeating with a different validation partition). This measure helps to assess how well our classifier generalises. \emph{Sensitivity} expresses the proportion of positives that are correctly classified as such; \emph{specificity} is the same but for negatives. Across all four measures, a higher percentage is better.

From our results, it is clear that the RBF-based classifier far outperforms the other two options. While RBF scores highly across all measures, the other classification functions lag far behind at around $60$ to $70\%$; they are much too simple for the datasets we are considering. Intuitively, we believe linear or polynomial classifiers are insufficient because readings of different sensors in \swat are correlated in complicated ways which are beyond the expressiveness of these kinds of classifiers. Given this outcome, we choose RBF as our classification function.

\subsection{RQ2: How large should the time interval in feature vectors be?}
\label{sec:how_large_time_interval}

Our second experiment assesses the effect on accuracy of using different time intervals in the feature vectors. As discussed before, a feature vector is of the form $(\pi, \pi')$ where $\pi$ denotes the water tank levels at a certain time and $\pi'$ denotes the levels after $d$ time units. Intuitively, using these feature vectors, the learnt model characterises the effects of mutants after $d$ time units. On the one hand, an abnormal system behaviour is more observable if this interval $d$ is larger (as the modified PLC control program has more time to take effect). On the other hand, having an interval that is too large runs the risk of reporting abnormal behaviours too late and thus potentially resulting in some safety violation.

Table~\ref{tab:comparison_interval} presents the results of a comparison of \emph{accuracy} and \emph{cross-validation accuracy} (both defined as for RQ1) across classifiers based on 100, 150, $\dots$ 300 ms time intervals. SVM-RBF was used as the classification function, and abnormal data was generated from 700 mutants.
 
 \begin{table*}[!t]
 	\renewcommand{\arraystretch}{1.3}
 	\caption{Effect of increasing the time interval on accuracy of SVN-RBF function}
 	\label{tab:comparison_interval}
 	\centering
 \begin{tabular}{c|c|c}
 \#time interval &   accuracy & cross-validation accuracy  \\
 \hline
  100&$90.98\%$  &$88.68\%$       \\ 
 \hline
 150 & $90.04\%$  & $90.01\%$       \\ 
 \hline
  200&$90.12\%$   &  $90.08\%$     \\ 
  \hline
  250& $91.05\%$  & $90.99\%$      \\ 
  \hline
  300&$90.05\%$   & $90.99\%$      \\ 
 \end{tabular}
 \end{table*}

 The results match the intuition mentioned earlier, although the accuracy stabilises much more quickly than we initially expected (at around 150ms time intervals). The time interval of 250ms has, very slightly, the best accuracy, so we continue to use it in the remaining experiments.

\subsection{RQ3: How many mutants do we need?}

Our third experiment assesses the effect on accuracy from using different numbers of mutant simulators to generate abnormal data. We are motivated to find the point at which accuracy stabilises, in order to avoid the unnecessary computational overhead associated with larger numbers of mutants.

\begin{table*}[!t]
	\renewcommand{\arraystretch}{1.3}
	\caption{Effect of increasing the number of mutants on accuracy of SVN-RBF function}
	\label{tab:comparison_mutants}
	\centering
\begin{tabular}{c|c|c|c}
\#mutants &\#effective mutants &   accuracy & cross-validation accuracy  \\
\hline
 300& 23&$63.01\%$   &$81.91\%$        \\ 
\hline
 400& 31 &$83.01\%$ & $89.01\%$      \\ 
\hline
 500& 62 &$90.07\%$   & $89.08\%$      \\ 
\hline
 600& 76 &$91.04\%$   & $90.89\%$      \\ 
\hline
 700& 91 &$91.05\%$   & $90.99\%$      \\ 
\end{tabular}
\end{table*}

Table~\ref{tab:comparison_mutants} presents a comparison of \emph{accuracy} and \emph{cross-validation accuracy} (both defined as for RQ1) across classifiers learnt from the data generated by 300, 400, 500, 600, and 700 \emph{mutants}. Our mutant sets are inclusive, i.e.~the set of 700 mutants includes all the mutants in the set of 600 in addition to 100 distinct ones. We also list how many of the generated mutants are \emph{effective}, in the sense that they can be compiled, run, and cause some abnormal physical effect with respect to at least one of the initial configurations. We used SVM-RBF as the classification function, collecting feature vectors (see Section~\ref{sec:collecting_feature_vectors}) with a time interval of 250ms.

The results indicate that both accuracy and cross-validation accuracy start to stabilise in the 90s from 500 mutants (62 effective mutants) onwards. It also shows that with fewer mutants (e.g.~300 mutants / 23 effective mutants) it is difficult to learn a classifier with acceptable accuracy. Given the results, we choose 600 as our standard number of mutants to generate.

 \subsection{RQ4: Is our model a physical invariant of the system?}\label{sec:eval_invariant}

Our fourth experiment is to establish whether or not there is statistical evidence supporting that the learnt model is a \emph{(physical) invariant} of the system, i.e.~it correctly classifies the data in normal traces as normal with accuracy greater or equal to some threshold $\theta$. We perform SMC as described in Section~\ref{sec:validating_classifier}, sampling positive feature vectors derived from a new and distinct initial configuration, setting the acceptable error bounds at a standard level of 0.05, and setting the threshold as $\theta = 91.04\%$ (i.e.~the accuracy of the classifier learnt from 600 mutants and a feature vector interval of 250ms). Our implementation performs hypothesis testing using SPRT, randomly sampling feature vectors and applying the classifier until SPRT's stopping criteria are met. If the sampled data is not enough, we sample additional feature vectors from the traces of additional new initial configurations.

Our SMC implementation repeated the overall steps above five times, each with normal data derived from a different distinct initial configuration (falling within normal operational ranges). In each run, our classifier passed, without requiring data to be sampled from traces of additional configurations. This provides some evidence that the classifier is an invariant of the \swat simulator. This is not surprising: in Section~\ref{sec:eval_classification_functions} we found that the sensitivity of the classifier was very high ($99.28\%$), i.e.~the proportion of positive feature vectors that it classified as such was very high. Our SMC implementation evaluates for the same property but seeks statistical evidence.

 \subsection{RQ5: Is our model useful for detecting attacks?}\label{sec:eval_usefulness}

Our final experiment assesses whether our learnt invariant is effective at detecting different kinds of attacks, i.e.~whether it classifies feature vectors as negative once an attack has been launched. First, we investigate network attacks, in which an attacker is assumed to be able to manipulate network packets containing sensor readings (read by PLCs) and signals (read by actuators). Second, we investigate code-modification attacks (i.e.~manipulations of the PLC programs), by randomly modifying the different PLC programs in the simulator and determining whether any resulting physical effects are detected. If able to detect the latter kind of attacks, the invariant can be seen as physically attesting the integrity of the PLC code.

\begin{table*}[!t]
	\renewcommand{\arraystretch}{1.3}
\caption{Results: detecting network attacks involving motorised valves (MV), pumps (P), and level indicator transmitters (LIT)}
\label{tab:results_network_attacks}
	\centering
\begin{tabular}{c|c|c|c|c|c}
attack \#  & attack point &  start state & attack   & detected  & accuracy\\
\hline
1 &MV‐101 &MV‐101 is closed &Open MV‐101  &yes &$89.67\%$\\
\hline
2 &P‐102 &P‐101 is on whereas P‐102 is off &Turn on P‐102  &yes  &$90.01\%$\\
\hline
3 &LIT‐101 &Water level between L and H &Increase by 1mm every second  &eventually &$63.11\%$\\
\hline
4 &LIT‐301 &Water level between L and H &Water level increased above HH  &yes&  $99.86\%$\\
\hline
5 &MV‐504 &MV‐504 is closed &Open MV‐504  &yes &$92.11\%$\\
\hline
6 &MV‐304 &MV‐304 is open &Close MV‐304  &yes&$88.01\%$ \\
\hline
7 &LIT‐301 &Water level between L and H &Decrease water level by 1mm each second  &eventually&$56.97\%$ \\
\hline
8 &MV‐304 &MV‐304 is open & Close MV‐304  &yes &$90.16\%$ \\
\hline
9 &LIT‐401 &Water level between L and H &Set LIT‐401 to less than L  &yes &$89.36\%$  \\
\hline
10 &LIT‐301 &Water level between L and H &Set LIT‐301 to above HH  &yes & $99.07\%$ \\
\hline
11 &LIT‐101 &Water level between L and H &Set LIT‐101 to above H  &yes & $91.12\%$ \\
\hline
12 &P‐101 &P‐101 is on &Turn P‐101 off  &yes &$92.06\%$ \\
\hline
13 &P‐101; P‐102 &P‐101 is on; P‐102 is off &Turn P‐101 off; keep P‐102 off  &yes &$91.62\%$ \\
\hline
14 &P‐302 &P302 is on &Close P‐302  &yes &$90.91\%$ \\
\hline
15 &LIT‐101 &Water level between L and H &Set LIT‐101 to less than LL  &yes &  $89.37\%$\\
\end{tabular}
\end{table*}

\myparagraph{Network attacks.} Table~\ref{tab:results_network_attacks} presents a list of network attacks that we implemented in the \swat simulator, and the results of our invariant's attempts at classifying them. Our attacks are from a benchmark of attacks that were performed on the \swat testbed for the purpose of data collection~\cite{Goh-et_al16a}. These attacks cover a variety of attack points, and were designed to comprehensively evaluate the robustness of \swat under different network attacks. Of the 36 attacks, we implemented the 15 that could be supported by the ODEs of (and thus had an effect on) the \swat simulator. The attacks are all achieved by (simulating) the manipulation of the communication taking place over the network, i.e.~hijacking data packets and changing sensor readings before they reach the PLC, and actuator signals before they reach the valves and pumps. The attacks cover a variety of \emph{attack points} in the \swat simulator: these are documented online~\cite{SWaT-Reference}, but intuitively represent motorised valves (MV), pumps (P), and level indicator transmitters (LIT). The table indicates whether or not the invariant was able to \emph{detect} each attack, and the \emph{accuracy} with which it labels the feature vectors (here, this reflects the percentage of feature vectors labelled as negative \emph{after} the attack has been launched). If the accuracy is high (above a threshold of $85\%$), we deem the attack to have been detected. Note that for attacks manipulating the sensor readings (LITs) read by PLCs, we assume that the \emph{correct} levels are logged by the historian.

As can be seen, all of the attacks were successfully detected. For all the attacks except \#3 and \#7, this is with very high accuracy (around~$90\%$ and above). This is likely because these attacks all trigger an immediate state change in an actuator (opening/closing a valve; switching on/off a pump), either by directly manipulating a control signal to it, or indirectly, by reporting an incorrect tank level and causing the PLC to send an inappropriate signal instead (e.g.~attack \#4 causes the PLC to switch on a pump to drain the tank, even though the water level is not actually high). Attacks \#3 and \#7 are not detected initially, hence the lower accuracy (approx.~$60\%$), because the sensor for the tank level is manipulated slowly, by 1mm per second. As a result, it takes more time to reach the threshold when the PLC opens a valve or switches on a pump, at which point the attack has a physical effect. If measuring from this moment onwards, Attack \#3 would have an accuracy of $99.83\%$ and \#7 an accuracy of $99.72\%$---hence our judgements of detected \emph{eventually}.

Overall, the results suggest that our invariant is successful at detecting network attacks when they lead to unusual physical behaviour, and thus might be useful in monitoring a system in combination with complementary defence mechanisms (e.g.~for ensuring the integrity of the communication links).

\myparagraph{Code modification attacks.} Table~\ref{tab:results_attestation} presents the results of some code modification attacks, and our invariant's ability to detect them. Unlike for network attacks, there is no benchmark of code modification attacks to use for \swat. In lieu of this, we randomly generated 40 \emph{effective mutants} (distinct from those in our learning phase), each consisting of a single mutation to a PLC program controlling some \emph{stage} of the \swat simulator. We generated data from these mutants with respect to our 20 initial configurations, collected feature vectors, and applied our invariant. The table reports how many of the mutants were \emph{detected} and with what \emph{accuracy} (we determine whether a feature vector should be positive or negative analogously to how we labelled feature vectors derived from mutant traces). After grouping the attacks with respect to the PLC program they affect, we report both the average accuracy for \emph{all} attacks as well as for only those that were \emph{detected}.

Our invariant was able to detect 32 of the 40 mutants. Upon manual investigation, we believe the reason it was unable to detect the remaining mutants was because they generated data traces that were too similar to the normal behaviour of the system. Similar to our network attacks, when a code modification attack led to an unexpected change in the states of valves and pumps, the attack was detected. The results suggest that the invariant could be effective for physically attesting the PLCs, i.e.~by monitoring the physical state of the system for any unexpected behaviours that could be caused by modified control code. Of course, an intelligent attacker may manipulate the code in a way that is not sufficiently captured by random modifications: seeking a more realistic attestation benchmark set is thus an important item of future work.

\begin{table*}[!t]
	\renewcommand{\arraystretch}{1.3}
\caption{Results: detecting code modification attacks}
\label{tab:results_attestation}
	\centering
\begin{tabular}{c|c|c|c|c}
attack stage  &   \# effective mutants  &  \# detected & accuracy (detected) & accuracy (all) \\
\hline
PLC 1 &8 &5 & $99.82\%$ & $71.54\%$\\
\hline
PLC 3 &20 &17 & $99.89\%$ &  $92.12\%$\\
\hline
PLC 4 &4 &4 & $99.29\%$ & $99.29\%$ \\
\hline
PLC 5 &5 &3 & $99.43\%$ & $81.20\%$ \\
\hline
PLC 6 &3 &3 & $99.87\%$ & $99.87\%$ \\
\hline
summary &40 & 32 & $99.84\%$ & $88.20\%$  \\

\end{tabular}
\end{table*}

\subsection{Threats to Validity}

Finally, we remark on some threats to the validity of our evaluation:

\begin{itemize}
	\item[(1)] Our dataset is limited to a single system: the \swat simulator;
	\item[(2)] Data traces were generated with respect to a fixed set of initial configurations;
	\item[(3)] We used randomly generated code modification attacks, rather than code modifications injected by an intelligent attacker.
\end{itemize}

Due to (1), it is possible that our results do no generalise to other CPSs. Because of (2), it is possible that normal but rarely occurring behaviours may have been missed in the training phase, and thus may be classified incorrectly by our invariant. These behaviours may also have been missed from the data traces used in the validation phase (SMC). Because of (3), it could be possible that our results do not apply to real code modification attacks designed by attackers with knowledge of the system.

\section{Related Work} \label{related_work}

Anomaly detection has been widely applied to \cps in order to detect unusual behaviours (e.g.~possible attacks) from their data~\cite{Cheng-Tian-Yao17a,Harada-et_al17a,Hofbaur-Williams02a,Hofbaur-Williams04a,Narasimhan-Biswas07a,Pasqualetti-Dorfler-Bullo11a,Teixeira-et_al12a,Verma-et_al04a,Zhao-et_al05a}. Many of these approaches, however, require prior knowledge about the internals of the system---our technique avoids this and attempts to construct a model systematically and automatically.

The idea of detecting attacks by monitoring physical invariants has been applied to a number of \cps~\cite{Choudhari-et_al13a,Paul-et_al14a}. Typically, however, the invariants are \emph{manually} derived using the laws of physics and domain-specific knowledge. Moreover, they are derived for specific, expected physical relationships, and may not capture other important patterns hiding in the sensor data. Manual invariants have also been derived for stages of the \swat testbed itself~\cite{Adepu-Mathur16b,Adepu-Mathur16a}.

Apart from monitoring physical invariants, the \swat testbed has also been used to evaluate other attack detection mechanisms, such as a hierarchical intrusion detection system for monitoring network traffic~\cite{Ghaeini-Tippenhauer16a}, and anomaly detection approaches based on unsupervised machine learning~\cite{Goh_et-al17a,Inoue-et_al17a}. The latter approaches were trained and evaluated using an attack log~\cite{Goh-et_al16a} from the testbed itself. As our approach was evaluated on the \swat simulator, an immediate and direct comparison with our results is not possible. However, we believe that our supervised approach would lead to higher sensitivity, and plan to do a proper comparison to confirm or refute this.

Mutations are applied by Brandl et al.~\cite{Brandl-Weiglhofer-Aichernig10a}, but to specifications of hybrid systems (rather than to the PLC programs themselves) in order to derive distinguishing model-based test cases that can be seen as classifiers. A discrete view of the system is used for generating test cases, with qualitative reasoning applied to represent the continuous part.

It is possible to obtain strong guarantees about the behaviour of a \cps by applying formal verification, but only with accurate enough models of the controllers and ODEs. With these, the \cps can be modelled as a hybrid system and a variety of established techniques can be applied (e.g.~model checking~\cite{Frehse-et_al11a}, SMT solving~\cite{Gao-Kong-Clarke13a}, non-standard analysis~\cite{Hasuo-Suenaga12a}, concolic testing~\cite{Kong-et_al16a}, runtime model validation~\cite{Mitsch-Platzer14a}, or theorem proving~\cite{Platzer-Quesel08a,Quesel-et_al16a}). With discretised models of the physical part, classical modelling and verification techniques can also be applied, e.g.~as demonstrated for some properties of \swat~\cite{Kang-et_al16a,Rocchetto-Tippenhauer17a}.

\section{Conclusion} \label{conclusion}

We proposed a novel approach for automatically constructing invariants of \cps, in which supervised ML is applied to traces of data obtained from \plc programs that have been systematically mutated. We implemented it for a simulator of the \swat raw water purification plant, presenting a framework that can generate large quantities of mutant \plc programs, data traces, and feature vectors. We used SVM-RBF to learn an expressive model, and validated it as characterising an invariant property of the system by applying cross-validation and statistical model checking. Finally, we subjected the simulator to 55 network and code modification attacks and found that the invariant was able to detect 47 of them (missing only 8 code modification attacks that had a limited effect on the water tank levels), suggesting its efficacy for monitoring attacks and physically attesting the \plcs at runtime.

Future work should seek to address the current complexity of the learnt invariants without reducing their effectiveness at detecting attacks, in order to bring them within reach of stronger validation approaches than SMC, e.g.~symbolic execution~\cite{Chen-Poskitt-Sun16a}. It should also seek to make the approach more practical for real \cps such as the \swat testbed (not just its simulator), by finding ways of reducing the amount of data that must be collected. One way we could achieve this is by applying mutations more effectively, reducing the amount of abnormal data we reject for being indistinguishable from normal traces. For example, we could use domain knowledge to focus the application of mutation operators to parts of the \plc code more likely to lead to useful abnormal traces. In future work we would also like to assess the generalisability of our approach by implementing it for other testbeds or simulators, especially those for applications other than water treatment. Finally, we would like to compare our supervised learning approach against some recently proposed unsupervised ones for \swat~\cite{Goh_et-al17a,Inoue-et_al17a}, in order to clarify whether or not the overhead of collecting abnormal data pays off in terms of the accuracy of the invariant and its ability to detect attacks.\\

\ifCLASSOPTIONcompsoc
  \section*{Acknowledgments}
\else
  \section*{Acknowledgment}
\fi

We thank Jingyi Wang for assisting us with statistical model checking, and are grateful to Sridhar Adepu and the anonymous referees for their helpful comments and criticisms. This work was supported in part by the National Research Foundation~(NRF), Prime Minister's Office, Singapore, under its National Cybersecurity R\&D Programme (Award No.~NRF2014NCR-NCR001-040) and administered by the National Cybersecurity R\&D Directorate.

\bibliographystyle{IEEEtran}

\bibliography{references}

\end{document}